\documentclass[
          ,draft            
  ]
  {aipproc}

\layoutstyle{6x9}

\usepackage{graphicx,psfrag}
\newcommand{\bda}{\begin{\displaymath}\begin{array}{rl}}
\newcommand{\eda}{\end{array}\end{displaymath}}
\newcommand{\be}{\begin{equation}}
\newcommand{\ee}{\end{equation}}
\newcommand{\bdm}{\begin{displaymath}}
\newcommand{\edm}{\end{displaymath}}
\newcommand{\bea}{\begin{eqnarray}}
\newcommand{\eea}{\end{eqnarray}}

\newcommand{\fs}{\; \; .}
\newcommand{\co}{\; \; ,}

\newcommand{\ubar}{\overline{\rule[0.42em]{0.4em}{0em}}\hspace{-0.5em}u}
\newcommand{\dbar}{\,\overline{\rule[0.65em]{0.4em}{0em}}\hspace{-0.6em}d}

\newcommand{\ChPT}{$\chi$PT\hspace{1mm}}

\begin{document}

\title{Model independent determination of the $\sigma$ pole}

\classification{11.30.Rd, 11.55.Fv, 11.80.Et, 12.39.Fe, 13.75.Lb}
\keywords      {QCD, dispersion theory, mesons}

\author{H.~Leutwyler}{
  address={Institute for Theoretical Physics, University of 
  Bern,\\ Sidlerstrasse 5, CH-3012 Bern, Switzerland}
email={leutwyler@itp.unibe.ch}
}

\begin{abstract}The first part of this report reviews recent developments at the interface between lattice work on QCD with light dynamical quarks, effective field theory and low energy precision experiments. Then I discuss how dispersion theory can be used to analyze the low energy structure of the $\pi\pi$ scattering amplitude in a model independent manner. This leads to an exact formula for the mass and width of the lowest few resonances, in terms of observable quantities. As an application, I consider the pole position of the $\sigma$, paying particular to error propagation in the numerical analysis. The report is based on work done in collaboration with Irinel Caprini and Gilberto Colangelo \cite{CCL}.  

\begin{center}Contribution to the proceedings of the Workshop on Scalar Mesons and Related Topics,\\ honouring the 70th birthday of Michael Scadron, Lisbon, Portugal, Feb.\ 11-16, 2008. \end{center}
\end{abstract}

\maketitle

\section{Motivation}
QCD with massless quarks is the ideal of a theory: it does not contain a
single dimensionless free parameter. At high energies, the degrees of freedom
occurring in the Lagrangian are suitable for a description of the phenomena
and the interaction among these degrees of freedom can be treated as a
perturbation. At low energies, on the other hand, QCD reveals a rich spectrum
of hadrons, the understanding of which is beyond the reach of perturbation
theory. In my opinion, one of the main challenges within the Standard Model is
to understand how an intrinsically simple beauty like QCD can give rise to the
amazing structures observed at low energy.

The progress achieved in understanding the low energy properties of QCD has
been very slow. A large fraction of the papers written in this field does not
concern QCD as such, but models that resemble it in one way or the other:
constituent quarks, NJL-model, linear $\sigma$ model, hidden local symmetry,
AdS/CFT and many others. Some of these may be viewed as simplified versions of
QCD that do catch some of the salient features of the theory at the
semi-quantitative level, but none provides a basis for a coherent
approximation scheme that would allow us, in principle, to solve QCD.
 
This talk concerns the model independent approach to the problem based on
dispersion theory. More precisely, I would like to show that the low energy
structure of the sector with the quantum numbers of the vacuum, $I=\ell=0$,
can quantitatively be understood on the basis of the symmetries of QCD,
without invoking any model, but instead relying on the available, rather crude
experimental information at energies above 1 GeV.

At low energies, the main characteristic of QCD is that the energy gap is very
small, $M_\pi\simeq $ 140 MeV. More than 10 years before the discovery of QCD,
Nambu \cite{Nambu} found out why that is so: the gap is small because the
strong interactions have an approximate chiral symmetry. Indeed, QCD does have
this property: for yet unknown reasons, two of the quarks happen to
be very light. The symmetry is not perfect, but nearly so: $m_u$ and $m_d$ are
tiny. The mass gap is small because the symmetry is ``hidden'' or
``spontaneously broken'': for dynamical reasons, the ground state of the
theory is not invariant under chiral rotations, not even approximately. The
spontaneuous breakdown of an exact Lie group symmetry gives rise to strictly
massless particles, ``Goldstone bosons''. In QCD, the pions play this role:
they would be strictly massless if $m_u$ and $m_d$ were zero, so that the
symmetry would be exact. The only term in the Lagrangian of QCD that is not
invariant under the group SU(2)$\times$SU(2) of chiral rotations
is the mass term of the two lightest quarks, $m_u\,\ubar u+m_d \,\dbar
d$. This term equips the pions with a mass. Although the
theoretical understanding of the ground state is still poor, we do have very
strong indirect evidence that Nambu's conjecture is right -- we know why the
energy gap of QCD is small.

\vspace{-0.3cm}
\section{Lattice results relevant for low energy QCD}
As pointed out by Gell-Mann, Oakes and Renner \cite{GMOR}, the square of the
pion mass is proportional to the strength of the symmetry breaking,
$M_\pi^2\propto (m_u+m_d)$. This property can now be checked on the lattice,
where -- in principle -- the quark masses can be varied at will. In
view of the fact that in these calculations, the quarks are treated
dynamically, the quality of the data is impressive. The masses are
sufficiently light for \ChPT to allow a meaningful extrapolation to the
quark mass values of physical interest. The results indicate that the ratio
$M_\pi^2/(m_u+m_d)$ is nearly constant out to values of $m_u, m_d$ that are
about an order of magnitude larger than in nature. 

The Gell-Mann-Oakes-Renner relation corresponds to the leading term in the
expansion in powers of the quark masses. At next-to-leading order, this
expansion contains a logarithm: 
\be\label{Mpi one loop} M_\pi^2=
M^2\left\{1 +\!\frac{M^2}{32\pi^2 F_\pi^2}\, \ln
  \frac{M^2}{\Lambda_3^2}\!+\!O(M^4)\right\}\co\hspace{1cm} M^2\equiv
B(m_u+m_d)\fs \ee 
Chiral symmetry fixes the coefficient of the logarithm in
terms of the pion decay constant $F_\pi$, but does not determine the scale
$\Lambda_3$. A crude estimate was obtained more than 20 years
ago \cite{GL}, on the basis of the SU(3) mass formulae for the pseudoscalar
octet. The result is indicated at the bottom of the left panel in Fig.1. 
\begin{figure}[thb] 
\parbox{15cm}{\psfrag{l3bar}{\hspace{-0.8cm}\raisebox{-0.5cm}
{$\displaystyle\bar{\ell}_3=\ln\frac{\Lambda_3^{\;2}}{M_\pi^2}$}}
\psfrag{l4bar}{\hspace{-1cm}\raisebox{-0.5cm}{$\displaystyle\bar{\ell}_4=\ln
    \frac{\Lambda_4^{\;2}}{M_\pi^2}$}} 
\includegraphics[height=.33\textheight,angle=-90]{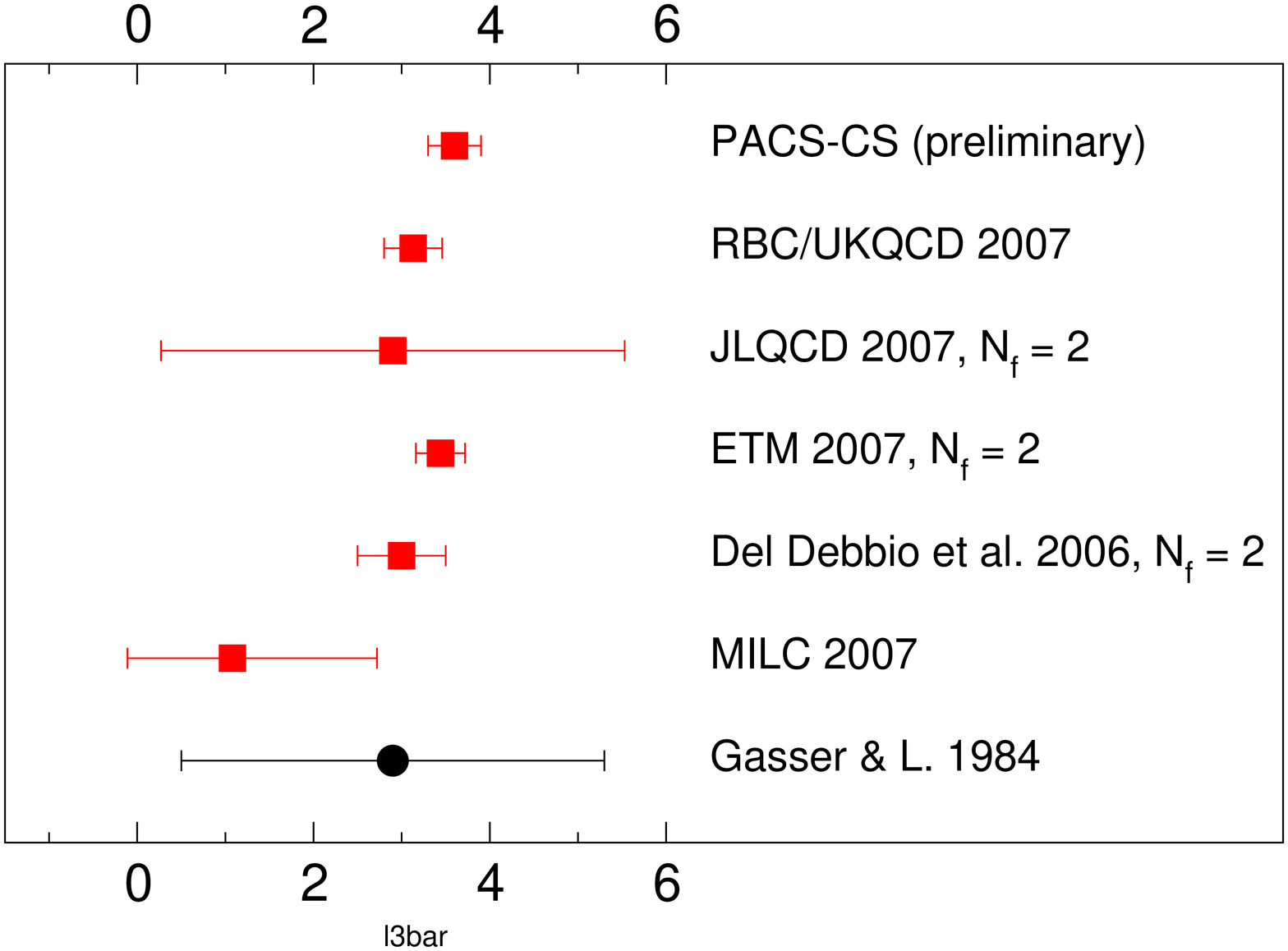}
\includegraphics[height=.33\textheight,angle=-90]{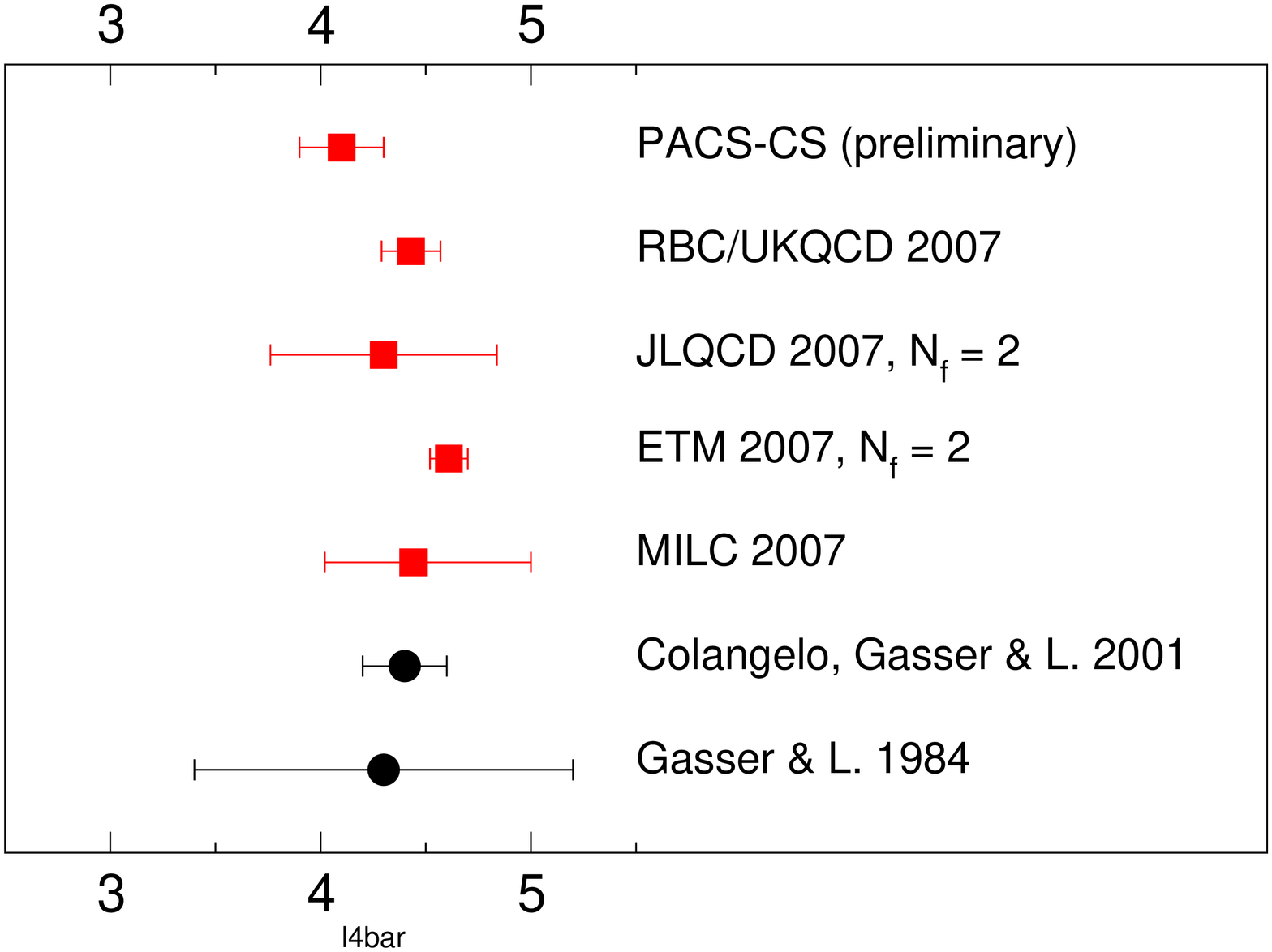}\\
\rule{0cm}{0.8cm}}

\caption{\label{fig:l3l4}Determinations of the effective coupling
  constants $\ell_3$ and $\ell_4$.}  
\end{figure}
The other entries represent recent lattice results for this quantity \cite{MILC}-\cite{PACS-CS}, which are considerably more accurate. The right panel shows the results for the coupling constant $\bar{\ell}_4$, which determines the quark mass dependence of the
pion decay constant at NLO of the chiral expansion. In that case, we obtained
a rather accurate result in 2001, from a dispersive analysis of the scalar
pion form factor \cite{CGL}. The lattice determinations of $\bar{\ell}_4$ have
reached comparable accuracy and corroborate the outcome of our analysis.

The hidden symmetry not only controls the size of the energy gap, but also
determines the interaction of the Goldstone bosons at low energies, among
themselves, as well as with other hadrons. In particular, as pointed out by
Weinberg \cite{Weinberg 1966}, the leading term in the chiral expansion of the
S-wave $\pi\pi$ scattering lengths (tree level of the effective theory) is
determined by the pion decay constant. The corresponding numerical values of
$a_0^0$ and $a_0^2$ are indicated by the leftmost dot in Fig.\ 2, while the
other two show the result obtained at NLO and NNLO of the chiral expansion,
respectively. The exotic scattering length $a_0^2$ is barely affected by
the higher order corrections, but the shift seen in $a_0^0$ is quite
substantial. The physics behind this enhancement of the perturbations
generated by $m_u$ and $m_d$ is well understood: it is a consequence of the
final state interaction, which is attractive in the $S^0$-wave, rapidly grows
with the energy and hence produces chiral logarithms with large coefficients. 

\begin{figure}[b] \parbox{15cm}{
    \includegraphics[height=.34\textheight,angle=-90]{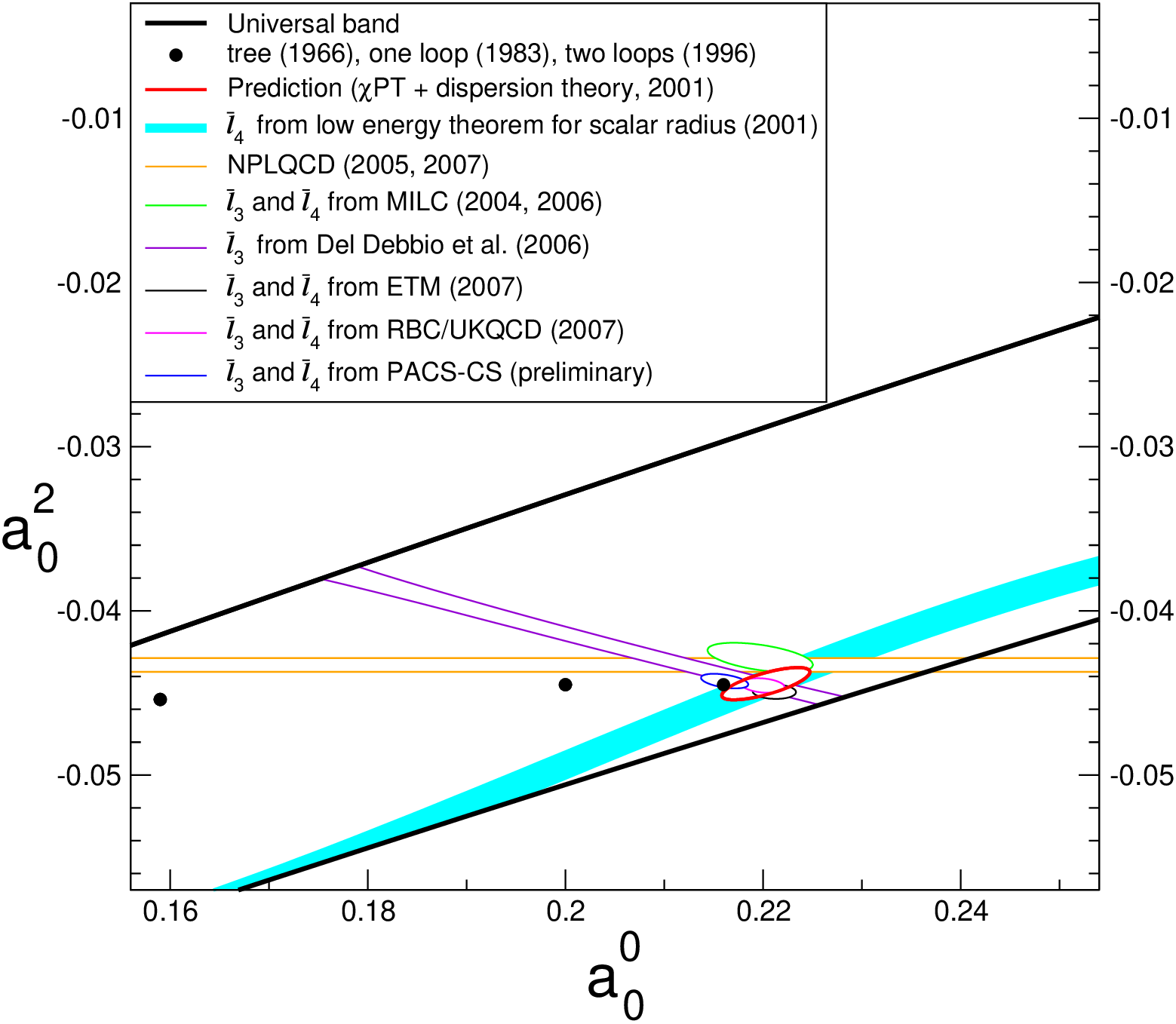}
    \includegraphics[height=.34\textheight,angle=-90]{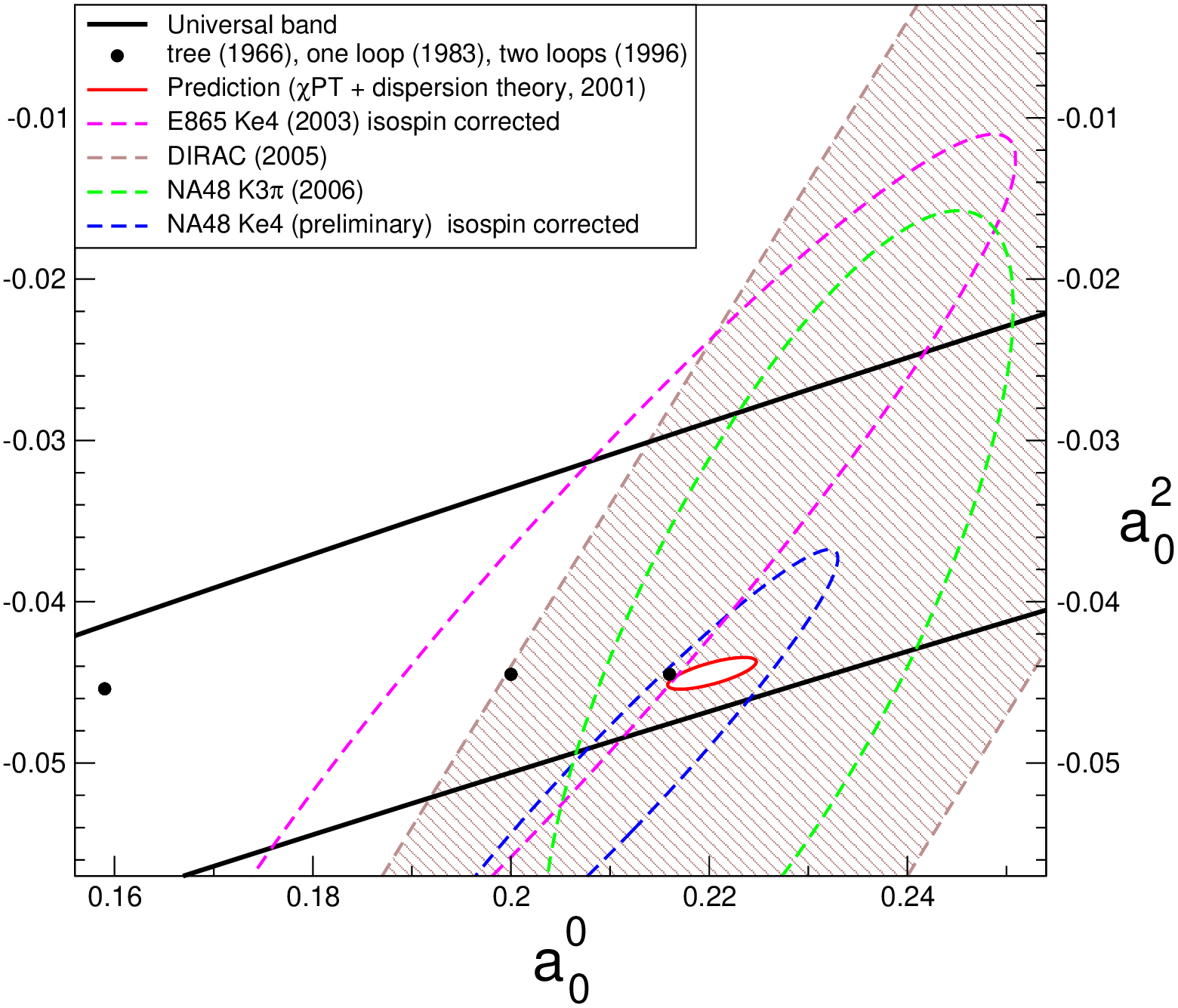}\\
    \rule{0cm}{0cm}}
\caption{Comparing the theoretical predictions for the $\pi\pi$ S-wave
  scattering lengths with lattice results (left) and with experiment (right).} 
\end{figure}
Near the center of the Mandelstam triangle, the
contributions from higher orders of the chiral expansion are small \cite{CGL}. Using
dispersion theory to reach the physical region, we arrived at the remarkably
sharp predictions for the two scattering lengths indicated on the left panel
of Fig.\ 2. Our analysis also shows that the corrections to Weinberg's low
energy theorem for $a_0^0, a_0^2$ are dominated by the effective coupling
constants $\bar{\ell}_3,\bar{\ell}_4$ discussed above -- if these are known,
the scattering lengths can be calculated within small uncertainties. Except
for the horizontal band, which represents a direct determination of $a_0^2$
based on the volume dependence of the levels \cite{NPLQCD}, all of the lattice results for the scattering lengths shown on the left panel of Fig.\ 2 are obtained in this
way from the corresponding results for $\ell_3$ and $\ell_4$. The figure
neatly demonstrates that the lattice results confirm the predictions for
$a_0^0,a_0^2$. 

\vspace{-0.3cm}
\section{Precision experiments at low energy}
The right panel of Fig.\ 2 compares the predictions for the scattering lengths
with recent experimental results. While the $K_{e4}$ data of E865 \cite{E865},
the DIRAC experiment\cite{DIRAC} and the NA48 data on the cusp in
$K\rightarrow 3\pi$ \cite{NA48 cusp} all confirm the theoretical expectations,
the most precise source of information, the beautiful $K_{e4}$ data of NA48
\cite{NA48 paper on Ke4}, gives rise to a puzzle. The Watson theorem implies
that -- if the electromagnetic interaction and the difference between $m_u$
and $m_d$ are neglected -- the relative phase of the form factors describing
the decay $K\rightarrow e\nu\pi\pi$ coincides with the difference
$\delta_0^0-\delta_1^1$ of scattering phase shifts. At the precision achieved,
the data on the form factor phase do not agree with the theoretical prediction
for the phase shifts. 

The origin of the discrepancy was identified by Colangelo, Gasser and Rusetsky
\cite{Colangelo Gasser Rusetsky}. The problem has to do with the fact that
neutral kaons may first decay into a pair of neutral pions, which then
undergoes scattering and winds up as a charged pair. The mass difference
between the charged and neutral pions affects this process in a pronounced
manner: it pushes the form factor phase up by about half a degree -- an
isospin breaking effect, due almost exclusively to the electromagnetic
interaction. Fig.\ 3 shows that the discrepancy disappears if the NA48 data
on the relative phase of the form factors are corrected for isospin breaking.
Accordingly, the range of scattering lengths allowed by these data, shown on
the right panel of Fig.\ 2, is in perfect agreement with the prediction.  The
intersection of this range with the band from the low energy theorem for the
scalar radius (left panel) yields $a_0^0=0.220(9)$.

In the mass range $M_{\pi\pi}>$ 350 MeV, Fig.\ 3 indicates a marginal
disagreement between NA48/2 and E865. While E865 collects all events in this
region in a single bin, the resolution of NA48/2 is better. 
\begin{figure}[thb]
\includegraphics[height=.32\textheight,angle=-90]{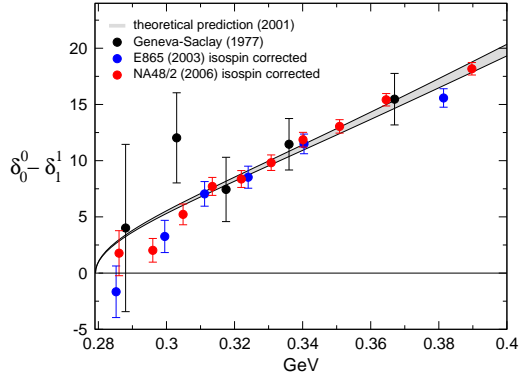}
\caption{Comparison of $K_{e4}$ data with the prediction for
  $\delta^0_0-\delta^1_1$}  
\end{figure} 
The fit to all $K_{e4}$ data is therefore dominated by NA48/2. For a detailed discussion of
these issues, I refer to the talks by B.\ Bloch-Devaux, G.\ Colangelo and J.\ 
Gasser at KAON 2007. I conclude that the puzzle is gone: $K_{e4}$ confirms the
theory to remarkable precision.

\vspace{-0.3cm}
\section{Roy equations}
As mentioned above, the straightforward evaluation of the chiral perturbation
series for the $\pi\pi$ scattering amplitude is useful only in a very limited
range of the kinematic variables -- definitely, the resonance poles are
outside this range. The domain of validity can be extended considerably by means of dispersion theory: analyticity, unitarity and crossing symmetry essentially determine the low energy properties of the scattering amplitude in terms of the two S-wave scattering lengths. I now wish to show that the properties of the lowest resonance of QCD can be worked out on this basis. 

From the point of view of dispersion theory, $\pi\pi$ scattering is
particularly simple: the $s$-, $t$- and $u$-channels represent the same
physical process.  As a consequence, the scattering amplitude
can be represented as a dispersion integral over the imaginary part and the
integral exclusively extends over the physical region \cite{Roy}. The
projection of the amplitude on the partial waves leads to a dispersive
representation for these, the Roy equations. I denote the S-matrix elements by $S^I_\ell=\eta^I_\ell \exp 2 i \delta^I_\ell$ and use the standard normalization for the corresponding partial wave amplitudes $t^I_\ell$:
\be\label{S}S_\ell^I(s)=1+2\,i\,\rho(s)\,t_\ell^I(s)\,,\hspace{2em}
\rho(s)=\sqrt{1-4M_\pi^2/s}\,.\ee The S-matrix elements and the partial wave amplitudes are analytic in the cut $s$-plane. There is a right hand cut ($4M_\pi^2<s<\infty$) as well as a left hand cut ($-\infty<s<0$). The Roy equation for the partial wave amplitdue with $I=\ell=0$ reads 
\bea\label{eq:Roy}
t_0^0(s)=a+(s-4M_\pi^2)b+\sum_{I=0}^2\,\sum_{\ell=0}^\infty\,
\int_{4M_\pi^2}^\infty ds'\, K^{0I}_{0\ell}(s,s')\,\mbox{Im}\hspace{0.1em}
t^{I}_\ell(s') .\eea
The equation contains two subtraction constants. As a consequence of crossing symmetry, these can be expressed in terms of the S-wave scattering lengths:
\be a = a_0^0, \hspace{1cm} b=(2a^0_0-5a^2_0)/12 M_\pi^2.\ee
The kernels $ K^{II'}_{\ell\ell'}(s,s')$ are explicitly known algebraic expressions which only involve the variables $s,s'$ and the mass of the pion. The integrals on the right hand side thus only involve observable quantities: the imaginary parts of the partial waves. 

As demonstrated by Roy, his equations rigorously follow from general principles of quantum field theory. They are valid for real values of $s$ in the interval $-4M_\pi^2<s<60M_\pi^2$ (the upper end is pushed up to $68M_\pi^2$ if the scattering amplitude obeys Mandelstam analyticity). Using known results of
general quantum field theory \cite{Martin book,Roy Wanders 1978}, we have shown that these equations also hold for complex values of $s$, in the intersection of the relevant Lehmann-Martin ellipses \cite{CCL}. 

The pioneering work on the physics of the Roy equations was carried out more
than 30 years ago \cite{MP}. The main problem encountered at that time was that
the two subtraction constants were not known. These dominate the dispersive representation at low energies, but since the data available at the time were consistent with a very broad range of $S$-wave scattering lengths, the Roy equation ana\-ly\-sis was not conclusive. 

The insights gained by means of \ChPT thoroughly changed
the situation. As discussed in detail in the first part of this report, the S-wave scattering lengths are now known very accurately. The main limitation in the numerical evaluation of equation (3) arises from the accuracy to which the imaginary parts can be pinned down. In this connection, it is essential that the Roy equations involve two subtractions, so that  the kernels fall off with the third power of the variable of integration. This ensures that the contributions from the low energy region dominate. Note that the left hand cut plays an important role here: taken by itself, the part of the kernel that accounts for the right hand cut falls off only with the first power of the variable of integration, but the high energy tail is cancelled by the contribution from the left hand cut. 

At low energies, the S- and P-waves dominate. In \cite{ACGL}, we solved the Roy equations for these waves only below 800 MeV, relying on the literature to estimate the contributions from higher energies and from the partial waves with $\ell \geq 2$. In the meantime, we have extended our analysis and now solve the Roy equations on their full range of validity, not only for the S- and P-waves, but also for the D- and F-waves. Treating the S-wave scattering lengths, the imaginary parts above $s_{max}=68 M_\pi^2$ and the elasticities as input, the Roy equations admit a two-parameter family of solutions. We identify the two free parameters with the values of the phase shifts $\delta^0_0$ and $\delta^1_1$ at 800 MeV. In view of the excellent experimental information about the vector form factor of the pion, the value of $\delta_1^1(800$ MeV) is reliably known, but the phenomenological information about $\delta^0_0(800$ MeV) is comparatively meagre -- this currently represents the main source of uncertainty in low energy $\pi\pi$ scattering and will be discussed in detail below.

\vspace{-0.3cm}
\section{Pole formula}

The positions of the poles represent universal properties of QCD, which are unambiguous even if the width of the resonance turns out to be large, but they concern the non-perturbative domain, where an analysis in terms of the local degrees of freedom -- quarks and gluons -- is not in sight. One of the reasons why the values for the pole position of the $\sigma$ quoted by the Particle Data Group cover a very broad range is that all but one of these either rely on models or on the extrapolation of simple parametrizations: the data are represented in terms of suitable functions on the real axis and the position of the pole is determined by continuing this representation into the
complex plane. If the width of the resonance is small, the ambiguities
inherent in the choice of the parametrization do not significantly affect the
result, but the width of the $\sigma$ is not small. For a thorough discussion of the sensitivity of the pole position to the freedom inherent in the choice of the parametrization, I refer to \cite{Caprini}.

The determination of the $\sigma$ pole provides a good illustration for the strength of the dispersive method and for the relative importance of the various terms on the right hand side of the Roy equations. The representation of the S-matrix element given above holds on a limited region of the first sheet. The pole sits on the second sheet, which is reached from the first by analytic continuation from the upper half plane into the lower half plane, crossing the real axis in the interval $4M_\pi^2<s< 16 M_\pi^2$, where the scattering is elastic. Now, unitarity implies that the values of the S-matrix element on the first and second sheets are related by $S^0_0(s)^{II}=1/ S^0_0(s)^{I}$. Hence a pole on the second sheet occurs if and only if $S^0_0(s)$ has a zero on the first sheet. Accordingly, we have an exact equation, which allows us to study the behaviour of the amplitude in the vicinity of the pole and find out whether -- in the limited region of the first sheet where the Roy equations are valid -- there are any resonances with the quantum numbers of the vacuum:  
\be S^0_0(s)=0.\ee 
As emphasized above, the representation of the S-matrix element that follows from the Roy equations exclusively involves observable quantities and can be evaluated for complex values of $s$ just as well as for real values -- for the above formula, an analytic continuation is not needed. 

Inserting our central representation for the scattering amplitude in (5), we find that, in the region where the Roy equations are valid, the function $S^0_0(s)$ has two zeros in the lower half of the first sheet: one at $\sqrt{s}= 441 - i\, 272$ MeV, the other in the vicinity of 1 GeV \cite{CCL}. While the first corresponds to the state $f_0(600)$, commonly referred to as the $\sigma$, the second zero represents the well-established resonance $f_0(980)$. Our analysis sheds little light on the properties of the latter, because the location of the zero is sensitive to the input used for the elasticity $\eta^0_0(s)$ -- the shape of the dip in $\eta^0_0(s)$ and the position of the zero represent two sides of the same coin.  For this reason, I only discuss the $\sigma$.

\vspace{-0.3cm}
\section{Discussion}

We are by no means the first to find a resonance in the vicinity of the above position. In the list of papers quoted by the Particle Data Group \cite{PDG 2007}, the earliest one with a pole in this ball park appeared more than 20 years ago \cite{Beveren}.  What is new is that we can perform a controlled error calculation, because our method is free of the systematic theoretical errors inherent in models and parametrizations. For this purpose, it is convenient to split the right hand side of the Roy equation for $t^0_0(s)$ into three parts:
\begin{enumerate}
\item Subtraction terms
\item Contribution from Im\hspace{0.1em}$t^0_0(s)$ below $K\bar{K}$ threshold
\item Contributions from higher energies and other partial waves
\end{enumerate}

{\bf ad 1.} As discussed above, the subtraction terms are determined by the S-wave scattering lengths. The prediction of \cite{CGL} reads $a_0^0=0.220\pm 0.005,\,a^2 _0=-0.0444\pm 0.0010$. Following the propagation of a change in $a_0^0$, we find that an increase by 0.005 shifts the pole position by $(-2.4 +i\,3.8)$ MeV, while the response to an increase in $a_0^2$ by 0.0010 is a shift of $(0.8 -i\, 4.0)$ MeV \cite{CCL}. 
\begin{figure}[thb]
\includegraphics[height=.3\textheight,angle=-90]{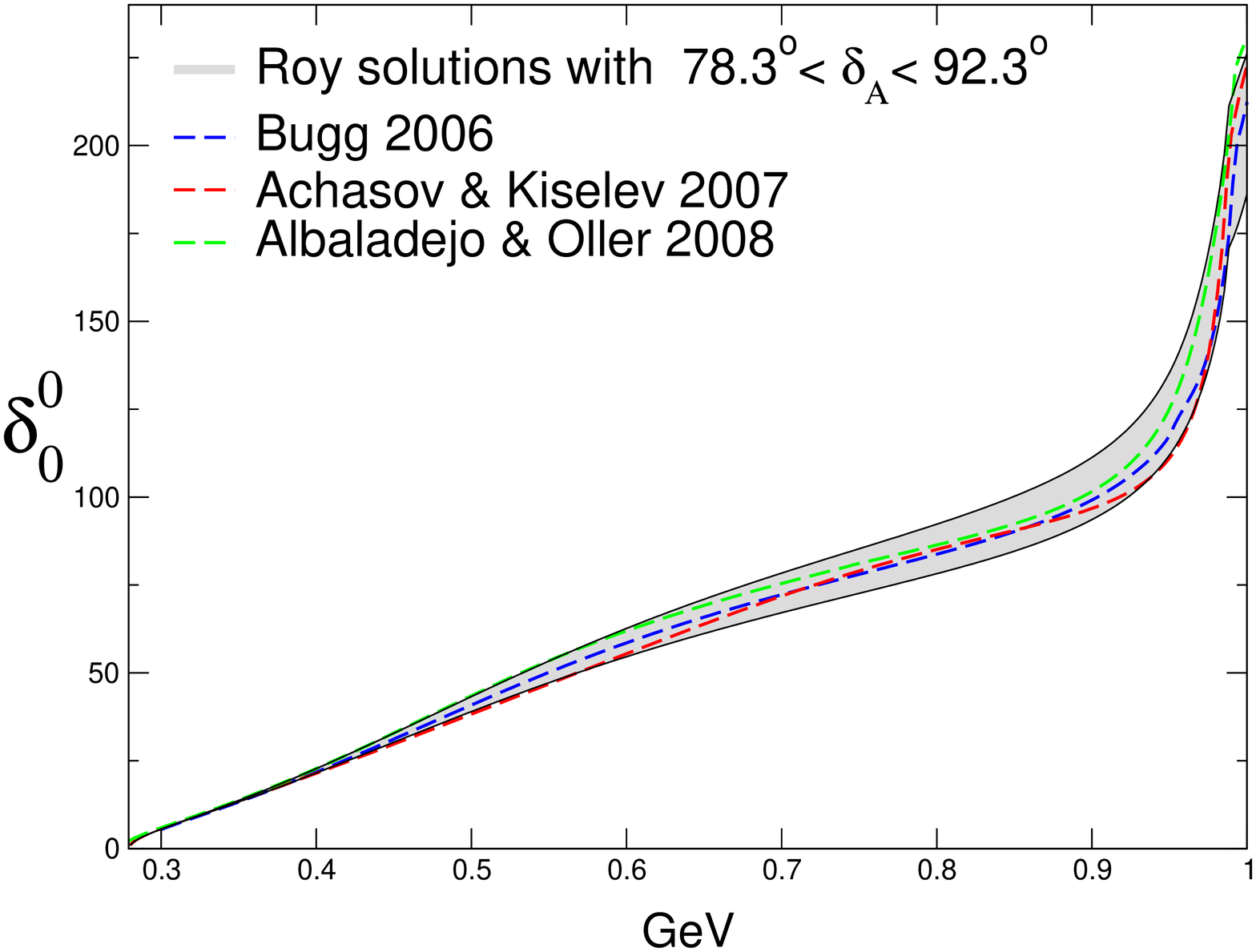}
\includegraphics[height=.3\textheight,angle=-90]{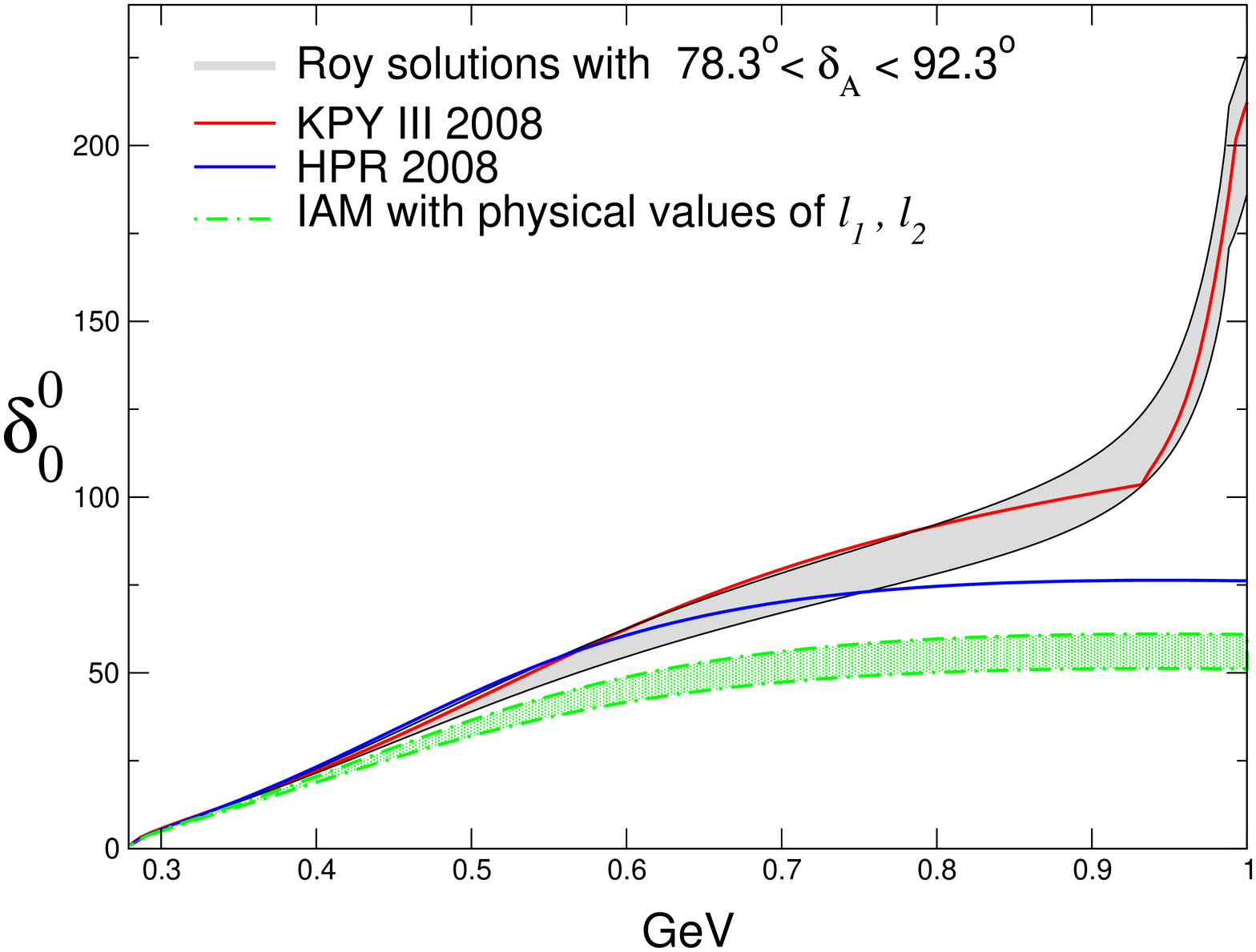}
\caption{\label{figdomain}
Behaviour of $\delta_0^0$ below $K\bar{K}$ threshold}  
\end{figure}
These numbers show that the error in the pole position due to the uncertainties in the subtraction constants are small.

{\bf ad 2. } Below $K\bar{K}$ threshold, the S-waves are elastic to a very good approximation. The function Im\hspace{0.1em}$t^0_0(s)$ shows a broad bump, nearly hits the unitarity limit somewhere between 800 and 900 MeV and then rapidly drops, because the phase steeply rises, reaching 180$^\circ$ in the vicinity of 2 $M_K$. Hence there is a pronounced dip in Im\hspace{0.1em}$t^0_0(s)$ near $K\bar{K}$ threshold: the behaviour of the imaginary part is controlled almost entirely by the phase shift $\delta^0_0(s)$. Fig.\ 4 shows several recent representations for this phase. Replacing the integral over our central representation for Im\hspace{0.1em}$t^0_0(s)$ from $4M_\pi^2$ to $4 M_K^2$ by the one of Bugg  \cite{Bugg} and leaving everything else as it is, the pole moves to 444 - i 267 MeV. Repeating the exercise with the representations of Achasov and Kiselev \cite{Achasov}, Albaladejo and Oller  \cite{Albaladejo} and Kami\'nski, Pel\'aez and Yndur\'ain \cite{KPYIII}, the pole is shifted to 438 - i 274 MeV, 451 - i 257 MeV and 458  - i 253 MeV, respectively.  As mentioned above, the solution of the Roy equation for $t^0_0(s)$ depends on the value of $\delta_A\equiv\delta^0_0($800 MeV). The shaded band in the figure shows the behaviour of the Roy solutions for $78.3^\circ<\delta_A<92.3^\circ$, the range used in \cite{CCL}. If the imaginary part of $t^0_0(s)$ is evaluated with the lower edge of this band, the pole occurs at 435 - i 276 MeV, while the upper edge corresponds to 456 - i 262 MeV.  

The representation used by Yndur\'ain and collaborators underwent a sequence of gradual im\-pro\-ve\-ments.  The most recent edition is definitely better than the earlier versions, but an important difference to our representation remains: as can be seen on the right panel of Fig.\ 4, the phase $\delta_0^0(s)$ of KPY III \cite{KPYIII} still contains a kink (discontinuity in the first derivative) at 932 MeV, as well as a hump below that energy. The kink arises because (i) two different parametrizations are used below and above 932 MeV and (ii) the correlations between the two regions imposed by causality are ignored. The deficiency is discussed in detail in \cite{Azores}, where it is shown that both the hump and the kink are artefacts, produced by the use of a parametrization that is not flexible enough. The analysis in \cite{Caprini} fully confirms this conclusion: among the 42 conformal parametrizations constructed there, all of those with an acceptable behaviour above 900 MeV run within the band of Roy solutions specified above. 

The hump is also responsible for the disagreement between \cite{CGL} and \cite{KPYIII} regarding the scattering lengths of the partial waves with angular momentum $\ell=1,2,3$ and the effective ranges with $\ell=0,1,2,3$: the contributions from Im\hspace{0.1em}$t^0_0(s)$  to the sum rules for these quantities are not the same (the sum rules are listed in Eqs. (14.1) and (14.3) of \cite{ACGL}). Within errors, the difference between these contributions reproduces the difference in the quoted results, without exception. 

I add a remark concerning the model of Hanhart, Pel\'aez and R\'ios \cite{HPR}, who apply the inverse amplitude method to improve the one loop approximation to the chiral perturbation series of SU(2)$\times$SU(2). In the original formulation of the model, the chiral expansion $t^0_0(s)=t_2(s)+t_4(s) + \ldots$ is unitarized with $t^0_0(s)=t_2(s)/\{1-t_4(s)/t_2(s)\}$, but this recipe fails in the vicinity of the Adler zero, because the term $t_4(s)$ does not vanish there. The deficiency is readily cured. It suffices to replace the IAM formula with
\be t^0_0(s)=\frac{\tilde{t}_2(s)}{1-\tilde{t}_4(s)/\tilde{t}_2(s)}\,,\hspace{1em} \tilde{t}_2(s)=t_2(s)-t_2(s_{A_4})\,,\hspace{1em}\tilde{t}_4(s)=t_4(s)+t_2(s_{A_4})\,,\ee
where $s_{A_4}$ is the position of the Adler zero in one loop approximation. Since $t_2(s_{A_4})$ represents a term of $O(p^4)$, the chiral expansion of (6) reproduces the one loop approximation of \ChPT\hspace{-0.25em}, also in the vicinity of the Adler zero. A similar recipe is used in \cite{HPR}. 

The model exclusively involves the coupling constants $F_\pi,\ell_1,\ldots,\ell_4$ of the effective Lagrangian. As discussed above, $\ell_3$ and $\ell_4$ are known quite well; $\ell_1$ and $\ell_2$ can be determined on phenomenological grounds \cite{CGL}. The result for the phase shift obtained by inserting the numerical values in the above formula is indicated on the right panel of Fig.\ 4. This shows that the model yields a decent approximation only below 500 MeV. The parametrization used by Hanhart eta al.\ \cite{HPR} is better, because these authors treat the coupling constants $\ell_1$ and $\ell_2$ as free parameters. This extends the range of energies where the IAM parametrization makes sense, but since the model does not account for the sharp increase in the phase towards $K\bar{K}$ threshold, it can at best give a semi-quantitative picture of the $\sigma$. For the parameter values adopted in \cite{HPR}, the zero of the denominator in (6) occurs at  444(6) - i 218(10) MeV: the mass is OK, but the width is too low by 100 MeV. Inserting the observed values of $\ell_1$ and $\ell_2$, the zero moves to  413(12) - i 269(12) MeV: now the width is OK, but the mass is too low.

{\bf ad 3.} Finally, I turn to the contributions of the third category: higher energies and other partial waves. Among these, the one from the P-wave, for example, is by no means negligible, but, as mentioned above, this wave is known very well. In fact, in the vicinity of the zero of $S^0_0(s)$, the sum of the contributions of this category can be worked out quite accurately. In \cite{CCL}, we estimated the net uncertainty in the pole position from this source at $\pm$ 4 $\pm$ i 6 MeV. As a check, we can simply replace our central representation for the contributions of category 3 by the one in \cite{KPYIII}, retaining our own representation only for the remainder. The operation shifts the pole position by - 0.6 - i 1.2 MeV, well within the estimated range.

\vspace{-0.3cm}
\section{Conclusion}
Adding the errors up in square, the result for the pole position becomes \cite{CCL} 
\be\label{eqmsigma} \sqrt{s_\sigma}=441\, \rule[-0.2em]{0em}{1em}^{+16}_{-\,8}-
\,i\;272\,\rule[-0.2em]{0em}{1em}^{+\,9}_{-12.5}\;\mbox{MeV}\,.\ee 
The error bars account for all sources of uncertainty and are an order of magnitude smaller than for the crude estimate $\sqrt{s_\sigma}$ = (400 - 1200) - i (250 - 500) MeV quoted by the Particle Data Group \cite{PDG 2007}. The dispersive representation of the S-matrix element also allows us to calculate the residue of the pole occurring on the second sheet, \be t^0_0(s)^{II}=\frac{r_\sigma}{s-s_\sigma}+\ldots\ee
Our preliminary result for the magnitude of the residue is $|r_\sigma|=0.218 \rule[-0.2em]{0em}{1em}^{+0.023}_{-\,0.010}$ GeV$^2$. 

\vspace{0.5em}
I thank Irinel Caprini and Gilberto Colangelo for a very pleasant collaboration and Chris Sachrajda, Daisuke Kadoh, Yoshinobu Kuramashi, David Bugg, Kolia Achasov, Lesha Kiselev, Miguel Albaladejo and Jos\'e Antonio Oller for informative discussions and correspondence. Also, it is a pleasure for me to thank George Rupp for the invitation and for warm hospitality during my stay at Lisbon.

\end{document}